# Chemical States and Local Structure in Cu-Deficient CuInSe$_2$ Thin Films: Insights into Engineering and Bandgap Narrowing

Ahmed Yousef Mohamed[a,b,*], Byoung Gun Han[a], Hyeonseo Jang[a], Jun Oh Jeon[a], Yejin Kim[a], Haeseong Jang[c], Min Gyu Kim[d], Kug-Seung Lee[d], and Deok-Yong Cho[a,*]

[a.] *Department of Physics, Jeonbuk National University, Jeonju 54896, Republic of Korea.*
[b.] *Department of Physics, Faculty of Science, South Valley University, Qena, 83523, Egypt.*
[c.] *Department of Advanced Materials Engineering, Chung-Ang University, Anseong 17546, Republic of Korea.*
[d.] *Pohang Accelerator Laboratory, Pohang 37673, Republic of Korea.*
*Corresponding authors: A. Y. Mohamed (yousef@jbnu.ac.kr) and D.-Y. Cho (zax@jbnu.ac.kr)

The Cu-deficient Cu$_x$InSe$_{\sim2}$ (x > 0.3) phase can be stabilized as a thin film. A uniform Cu-deficient composition with a chalcopyrite structure was obtained by the precision engineering of a two-step synthesis process involving electron-beam evaporation and Se vapor deposition. Detailed structural and chemical analyses were performed employing various X-ray and microscopic techniques to demonstrate that the chemical states and local structure in the Cu-Se-In tetrahedral networks change with the loss of Cu, the In–Se bond becomes shorter, and the In ions become excessively oxidized without phase separation. Moreover, the results indicate that the bandgap narrowing is primarily attributed to the reconstruction of In$_{3+\delta}$ 5s orbital states. The bandgap narrows from 1.51 eV to 1.4 eV, which is optimal for the photon absorber. Therefore, cation-deficient selenide is promising for stable nontoxic photovoltaics with tunable bandgaps.

**Introduction**

Copper indium selenide (CuInSe$_2$) has drawn considerable attention for its feasible applications in photovoltaic devices, such as photon absorbers in solar cells [1-4], light-emitting diodes [5-8], and in biological and catalysis applications [9-12]. Research on selenide compounds is particularly aimed at securing the controllability of optoelectronic properties with stable chemistry to eventually replace semiconductors that contain toxic elements such as cadmium or lead [13-15]. However, because ternary compounds bear inherent complexity in their structure and defect chemistry [16, 17], a full understanding of the role of intrinsic/extrinsic defects in fundamental science is yet to be achieved.

To date, various synthesis methods have been applied to produce compositions with optimal optoelectronic properties, including evaporation in vacuum [1, 18], thermal evaporation [6], molecular beam epitaxy [19], sputtering [20, 21], electrodeposition [22, 23], and electron beam evaporation [24-26]. The resultant compounds are often (particularly in the case of bulk synthesis) in the form of solid solutions in which more than one phase with different atomic concentrations is involved [27, 28]. Such a multiphase configuration hinders the straightforward assignment of the roles of each single phase, and it is desirable to synthesize the CuInSe$_2$ system (hereafter, abbreviated as CISe) in a single crystalline phase.

For this purpose, the electron beam (e-beam) evaporation of high-melting-point metals (Cu or In) can offer advantages because the pure metal particulates physically detached by the e-beam are deposited onto the substrate being highly focused and because the metal concentration ratios can be controlled very well. However, Se can vaporize at relatively low temperatures (< 700 °C) and easily penetrate the specimens. Therefore, a combination of the two deposition techniques (e-beam evaporation for Cu/In and vaporization for Se) would be ideal for the precise control of compositions.

The optoelectronic properties of CISe depend significantly on their composition. Earlier reports show that nonstoichiometric CISe suffers degradation of its optoelectronic properties [25, 29]. In particular, the concentration ratio of Cu versus In, i.e., $x$ = [Cu]/[In] in Cu$_x$InSe$_{\sim 2}$, appears to be a critical factor determining the functionality: Cu-rich CISe ($x$ > 1) would have higher carrier mobility than the stoichiometric CISe, while Cu-deficient CISe ($x$ < 1) suffers a weaker (better) recombination effect at the interface and more importantly, has a larger bandgap than the bulk stoichiometric CISe (~1.05 eV) [16, 20, 22, 23, 29, 30]. The underlying mechanism of optical photoabsorption under Cu-deficient conditions is yet to be elucidated, and the optical transition in Cu-deficient CISe could be attributed to the presence of Cu-vacancy-related defect states or to self-trapped excitonic states [17, 30-32]. In the former case, the weakened contribution of Cu 3d (owing to the loss of Cu) to the Se 4p–Cu 3d interband interaction might lower the valence band energy and increase the bandgap [17, 33]. However, in the latter case, the formation of excitonic shallow acceptor levels due to Cu vacancies could decrease the bandgap [16, 17, 28, 34].

Therefore, this work focused on demonstrating the correlation between Cu deficiency and the chemical, structural, and optoelectronic properties under almost controlled conditions. All the growth conditions were set identically except for the Cu amount, and a single chalcopyrite structure was maintained for all the compositions (0.34 ≤ $x$ ≤ 1.00) without any phase separation into, e.g., Cu$_2$Se + In$_2$Se$_3$. Several microscopic and spectroscopic techniques, including X-ray diffraction (XRD), scanning electron microscopy (SEM), transmission electron microscopy (TEM), synchrotron X-ray absorption spectroscopy (XAS), X-ray photoelectron spectroscopy (XPS), and ultraviolet-visible-near infrared (UV/Vis/NIR) optical spectrometry, were performed in parallel to verify the findings. All the results consistently indicate a systematic evolution in the In chemistry, Cu-Se-In local structure, and bandgap in accordance with the Cu concentration, suggesting the feasibility of bandgap tuning by cation deficiency within the chalcopyrite framework.

**Experimental details**

**Materials.** The following chemicals were purchased from Alfa Aesar (Thermo Fisher Scientific) and used: In powder (-100 mesh, 99.99%), Cu rod (0.08 inch, 99.99%), and Se powder (-200 mesh, 99.999%). In powder and a Cu rod were used as sources in the e-beam evaporation in vacuum, whereas Se powder was used as the source in the vaporization step after e-beam evaporation. A 5 mm-thick p-type Si-wafer (Silicon Materials Inc.) with native $SiO_2$ was used as the substrate.

**Samples preparation.** $Cu_xInSe_{\sim 2}$ thin films with various Cu concentrations were grown by a two-step process consisting of e-beam evaporation of In and Cu, followed by Se vapor deposition. A schematic of this process is illustrated in Fig. 1. In the first step, ~14 nm-thick In layer and Cu layers with different thicknesses (approximately 2, 3, 5, and 6 nm) were consecutively deposited on the $SiO_2$/Si wafer using an EFM3 e-beam evaporator (FOCUS GmbH). The thickness was estimated from the evaporation time after calibration using a thickness monitor. The rate of deposition was maintained at 0.5 Å/min for In and 0.1 Å/min for Cu, respectively. The base pressure of the vacuum chamber was $1 \times 10^{-8}$ torr.

In the second step, the deposited Cu/In layers (on $SiO_2$/Si) were exposed to Se vapor in a tube furnace. The specimens were placed in the hot center of a tube, and the Se powder was placed in a crucible vaporizer. The air inside the tube was first pumped out to a base pressure of $1 \times 10^{-3}$ torr and then filled with Ar gas several times to flush out the residue oxygen before starting the selenization to minimize unwanted oxidation. The furnace was heated slowly to reach the reaction temperature of 500 $^o$C (within 20 minutes), maintained at the temperature for 5 min, and then naturally cooled. During the heating process, In melts and In and Cu layers are mixed. At the same time, Se vapor permeates into the mixed Cu+In film and then ionized. However, the amount of Se ions in the CuInSe film is likely self-limiting in that it does not exceed the stoichiometric value of 2 in $CuInSe_2$ (as is confirmed by X-ray photoelectron spectroscopy). Consideration on the ionic diffusion processes is demonstrated in the Supplementary Information. After the selenization, the film swelled up to approximately 30–60 nm because of the inclusion of Se, and all the atoms in the film were mixed to form an almost uniformly thick layer.

We prepared five samples: four $Cu_xInSe_{\sim 2}$ films with different [Cu]/[In] ratios and a reference $In_2Se_3$ film (30 nm). The CISe samples are named CISe-1, CISe-2, CISe-3, and CISe-4 in the order of increasing Cu concentrations.

**Characterizations.** The crystal structure of the samples was examined using a multi-function X-ray Diffractometer (XRD, Rigaku Co., Japan) with monochromatic Cu K$\alpha_1$ radiation at the Korea Basic

Science Institute (KBSI) Daegu Center, Korea. XRD patterns were analyzed using the X'Pert HighScore Plus program.

The surface morphologies of the samples were examined using field-emission scanning electron microscopy (FE-SEM, SUPRA 40 VP, Carl Zeiss). The energy of the incident electron beam was set to 10 keV to enhance signals from the surface. Additionally, atomic distribution maps were obtained using the energy-dispersive X-ray (EDX) analysis supported by FE-SEM. Cross-sectional images of the CISe thin films were obtained using aberration-corrected transmission electron microscopy (Cs-TEM, JEM-ARM200F, JEOL Ltd.) with an acceleration voltage of 200 kV and high-angle annular dark-field scanning TEM (HAADF-STEM) with EDX mapping. The FE-SEM and Cs-TEM analyses were performed at the Center for University-Wide Research Facilities at Jeonbuk National University.

The [Cu]/[In] atomic concentration ratios in the $CuInSe_2$ chalcopyrite phase were determined using EDX analysis (from both FE-SEM and Cs-TEM). They were estimated to be 0.34, 0.46, 0.78, and 0.99 for CISe-1, CISe-2, CISe-3, and CISe-4, respectively (see SI†, Section 1 for more details).

The chemical environments around the In, Se, and Cu atoms were examined using X-ray absorption spectroscopy (XAS). Hard XAS at the Se K-, Cu K-, and In K-edges was performed at the 8C and 10C beamlines at the Pohang Light Source (PLS), Korea, in the fluorescence yield mode. The probing depth of the hard XAS exceeds 1 micron, so it reflects the chemistry throughout the film depth. The chemical states at the sample surface were identified using an XPS (Nexsa XPS system, Thermo Fisher Scientific) equipped with a monochromatic Al Kα source (1486.6 eV) at the KBSI Jeonju Center, Korea.

Optical measurements were performed using a UV/Vis/NIR optical spectrometer (Hitachi, Tokyo, Japan) at the KBSI Daegu Center, Korea. The spectrum of the Si substrate was subtracted from absorption spectra to eliminate the contributions from the substrate. The bandgap energies of $Cu_xInSe_{\sim 2}$ thin films were determined using Tauc plots.

**Results**

**XRD.** In Fig. 2a, the XRD patterns of $Cu_xInSe_{\sim 2}$ with various Cu concentrations are shown. For comparison, the XRD pattern of the reference $In_2Se_3$ film is also provided. The overall low thickness of the samples and high substrate background resulted in high noise levels in the XRD patterns. However, the $Cu_xInSe_{\sim 2}$ films exhibited clear XRD patterns (at 2θ = 26.6°, 44.2°, and 52.3°, corresponding to the (112), (220), and (312) reflections, respectively) that are consistent with a $CuInSe_2$ chalcopyrite structure (space group: $I\bar{4}2d$, ICDD:040-1487) [35]. The XRD results clearly show that the $CuInSe_2$ chalcopyrite structures are maintained in all the samples. This demonstrates the robustness of the crystal structures of Cu-deficient CISe, for example, CISe-1, synthesized by the combined e-beam + selenization process. This suggests that the thin films did not undergo phase separation into $Cu_2Se$ + $In_2Se_3$ based on the [Cu]/[In] ratio. Hence, the chemical, structural, and optoelectronic properties of the CISe specimens cannot be assimilated with a simple mixture of those in segregated end-members such as $Cu_2Se$ + $In_2Se_3$.

Meanwhile, the diffraction patterns of the reference sample (black curve) can be ascribed to a hexagonal $In_2Se_3$ phase (space group: $P6_1$, ICDD:00-040-1407) with the angles of 2θ = 25.0°, 27.5°, 37.5°, 38.9°,

39.8°, 44.0°, and 52.7° corresponding to (110), (006), (116), (211), (212), (300), and (306) reflections, respectively [36].

Figure 2a clearly shows that the overall peak intensities increase with increasing Cu concentration, implying improved crystallinity. The full-width-at-half-maximum (FWHM) values of the (112), (220), and (312) profiles decreased, indicating increasing crystallite sizes as the Cu concentration increased toward stoichiometry. The average crystallite sizes were calculated from the FWHM values of the peak profiles using the Debye-Scherrer method.

$$D = K\lambda/(\Gamma \cos\theta) ,  \quad (1)$$

where $\lambda$ is the X-ray wavelength (Cu K$\alpha_1$, 1.5406 Å), K is a shape factor set as 0.9, $\theta$ is the angle of the peak, and $\Gamma$ is the FWHM [37, 38]. The crystallite sizes estimated by using the average of the most intense peaks (i.e., (112), (220), and (312)), are 10.4 nm, 11.6 nm, 13.1 nm, and 15.7 nm for CISe-1, CISe-2, CISe-3, and CISe-4, respectively.

The deduced lattice constants $a$ (= $b$) and $c$ of the Cu$_x$InSe$_{\sim 2}$ thin films are summarized in Table 1. The lattice constants in a tetragonal structure ($I\bar{4}2d$) can be obtained from the interlayer spacing d$_{hkl}$ by the following equation (2) [39]:

$$1/(d_{hkl}^2) = (h^2 + k^2)/a^2 + l^2/c^2 , \quad (2)$$

**Table 1**. The compositions, lattice constants, and deduced average crystallite sizes in the Cu$_x$InSe$_{\sim 2}$ films.

| Sample | [Cu]/[In][a] | Lattice constants (Å) | | Crystallite size (nm) |
| --- | --- | --- | --- | --- |
| | | a = b | c | |
| CISe-1 | 0.34 | 5.76 | 11.63 | 10.4 |
| CISe-2 | 0.46 | 5.78 | 11.59 | 11.6 |
| CISe-3 | 0.78 | 5.79 | 11.61 | 13.1 |
| CISe-4 | 0.99 | 5.79 | 11.59 | 15.7 |

[a] See Supplementary Information†, Section 1, for details on the concentration ratios.

For all the samples, the lattice constants are similar to those in reference card no. 040-1487 in the ICDD database ($a$ = $b$ = 5.78 Å, $c$ = 11.62 Å) [35], which implies the evolution in the long-range structure upon composition changes is minimal.

**FE-SEM.** FE-SEM was performed to investigate the surface morphology and grain size of the CISe thin films. Figure 2b-e shows the FE-SEM images of the $Cu_xInSe_{\sim 2}$ thin films with different Cu concentrations. The average grain sizes in the FE-SEM images were estimated to be 44.5 nm, 52.3 nm, 59.0 nm, and 71.1 nm for CISe-1, CISe-2, CISe-3, and CISe-4, respectively, indicating a gradual increase as the Cu concentration increases.

The uniformity of the atomic distributions and concentration ratios in the $Cu_xInSe_{\sim 2}$ thin films were evaluated using EDX mapping analysis, as shown in Fig. S1†. The results of the EDX analysis indicate that Cu, In, and Se atoms are uniformly distributed over the thin film surface, which is consistent with the XRD findings of single $CuInSe_2$ phase formation throughout the films. Hence, the total intensity of each atomic species in the EDX data was used to obtain the average atomic concentration of each element throughout the film. See Table S1† in Supplementary Information for further details.

**Cs-TEM.** Cs-corrected TEM was used to obtain the cross-sectional images of the $Cu_xInSe_{\sim 2}$ thin films, as shown in Figs. 2f-i. Elemental analyses of Cu, In, and Se were performed via EDX mapping across the film/substrate interfaces in the four $Cu_xInSe_{\sim 2}$ samples (Fig. S2†). The total thickness of the samples increased as the [Cu]/[In] ratio increased within the range of 30–60 nm because of the addition of Cu. The cross-sectional TEM images revealed the grain structures of the samples across the film structure. The size of the $CuInSe_2$ grains increased progressively from CISe-1 (Cu-deficient) to CISe-4 (almost fully stoichiometric), as shown in Figs. 2f and 2i, respectively, which is consistent with the FE-SEM results (Figs. 2b-e). The smaller grains in the Cu-deficient samples imply a larger volume of grain boundaries, which might act as a source of imperfections in crystallinity or smaller crystallite sizes as shown in the XRD data.

The enlarged views (Figs. 2j,k) for CISe-1 and CISe-4, show that the d-spacings of some representative areas denoted in the figures were 3.358 Å and 2.05 Å, which correspond to the (112) and (220) reflections of a tetragonal $CuInSe_2$ phase ($I\bar{4}2d$, ICDD:040-1487) [35], respectively. The Fourier transformed diffraction images taken from the cross-sectional region (insets of Figs. 2j,k), show discrete patterns that can be primarily indexed to (112), (220), and (312) of $CuInSe_2$ chalcopyrite structure, together with faint concentric ring patterns typical for polycrystalline specimens. The clear spot-like patterns with regular displacements indicate the phase purity of both CISe-1 and CISe-4 without a considerable phase separation e.g. into $Cu_2Se+In_2Se_3$. This is consistent with the finding in the XRD data.

**XPS.** XPS was used to analyze the chemical states of each element at the surface. The XPS survey spectrum (Fig. S3†) of the $Cu_xInSe_{\sim 2}$ samples revealed the presence of Cu, In, Se, O, and C at the surface. All XPS spectra were calibrated to align the C 1s peak center at a binding energy of 284.6 eV. Figure 3 shows the (a) Cu 2p, (b) In 3d, and (c) Se 3d core-level spectra of the four CISe samples and reference $In_2Se_3$. All ionic species were in the expected oxidation states of $Cu^{1+}$, $In^{3+}$, and $Se^{2-}$ in the $CuInSe_2$ structure [40]. In the Se 3d XPS spectra, contributions from contaminated surfaces were also observed. The survey scans are shown in Fig. S3† in the Supplementary Information.

In Fig. 3a, the XPS spectra of Cu 2p showed two-peak features for spin-orbit-split Cu $2p_{3/2}$ and $2p_{1/2}$, with binding energies (BE) of approximately 932.0 eV and 951.9 eV, respectively. The binding energies and peak splitting value of 20.1 eV are consistent with those of monovalent Cu ($Cu^+$) [40-42]. In the case of In 3d, the features of $3d_{5/2}$ and $3d_{3/2}$ doublets were observed at BEs of ~444.6 eV and ~452.1 eV, respectively, corresponding to trivalent In ($In^{3+}$) [42-44]. Similarly, the presence of the $Se^{2-}$ state is evidenced by the Se 3d peaks at a BE of 54.6 eV for all samples, whereas the broad features at BE ~56 eV and ~59 eV correspond to oxidized Se due to air contamination at the surface [42, 45, 46].

As shown in Fig. 3c, the main Se 3d peaks (apart from the contributions of surface oxidation) hardly shift as the Cu concentration changes. In contrast, in Figs. 3a and 3b, the Cu $2p_{3/2}$ peaks and the In $3d_{5/2}$ peak shift systematically to higher energies, by ~0.1 eV and ~0.2 eV, respectively, as the [Cu]/[In] ratio decreases in the $Cu_xInSe_{\sim2}$ thin films. Notably, the In $3d_{5/2}$ peak became closer to the peak for $In_2Se_3$. The insets in Figs. 3a and 3b show this shift more clearly. The systematic shift indicates an increase in the Cu and In valences (although the nominal In valence in $CuInSe_2$ is the same as that in $In_2Se_3$) in accordance with the lack of Cu in the $CuInSe_2$ framework. Therefore, we can infer from the XPS data that the $Cu^{1+}$ and $In^{3+}$ ions at the surface of Cu-deficient CISe undergo slightly additional oxidation compared to stoichiometric $CuInSe_2$. The shift in BE for In is higher than that of Cu by ~0.1 eV, indicating that the $Se^{2-}$ ions in Cu-deficient CISe are more likely to bond with $In^{3+}$ cations and tend to attract more electrons from $In^{3+}$ [16, 17, 28], which is natural in the loss of Cu, while nevertheless maintaining the chalcopyrite structure. The relevant structural evolution will be scrutinized in the following sections.

From XPS analysis, we inferred that Cu deficiency resulted in a change in the bonding environment of the $In^{3+}$ ions. The detailed evolution of the bonding environment was not observed in the XRD of the long-range crystal structure. However, this may be observed in the local structure.

**Cu K-edge XAS.** Figure 4 shows the Cu K-edge X-ray absorption near edge structure (XANES) spectra and the first derivatives of the $Cu_xInSe_{\sim2}$ thin films with various concentrations of Cu. All of the samples exhibit the same absorption edge energy, and none of them have pre-edge peaks in the range of 8975–8980 eV, which is typically associated with the 1s→3d transition of $Cu^{2+}$ species [19, 47, 48]. The absence of pre-edge peaks was further confirmed by analyzing the first derivative of the Cu K-edge XANES spectra, as shown in Fig. 4b. This suggests that the valence of Cu atoms in the $Cu_xInSe_{\sim2}$ samples remain monovalent throughout the film depth, as in fully stoichiometric $CuInSe_2$, and it is not influenced by the concentration of Cu. However, at the surface, the Cu 2p XPS data showed a slight additional oxidation of $Cu^{1+}$, as shown in Fig. 3a.

The inset of Fig. 4a shows the Fourier transform (FT) extended X-ray absorption fine structure (EXAFS) spectra of the $Cu_xInSe_{\sim2}$ samples. The peak that appears at a radial distance (R) before the phase-shift correction around 2.0 Å for all the samples can be attributed to the Cu-Se bonds. The FT magnitudes and R values of the peaks were almost the same. To obtain the structural parameters of the Cu-Se coordination shells, we fitted the first-shell peaks at 1–3 Å in the inset of Fig. 4a for the Cu-Se single scatterings using ARTEMIS [49]. A passive electron reduction factor ($S_0^2$) of 0.7 was assumed for the fitting [50]. Table S2† lists the results of the EXAFS analyses, revealing that the coordination number N of the $Cu_xInSe_{\sim2}$ films remains constant at approximately 3.9. In contrast, as the concentration of Cu increases, the length of the Cu-Se bond slightly decreases from 2.421 to 2.407 Å. This finding is consistent with the calculated bond length of Cu-Se [16].

**Se K-edge XAS.** The Se K-edge XANES spectra of the $Cu_xInSe_{\sim2}$ films and $In_2Se_3$ reference sample are presented in Figs. 4c and 4d. Two main peaks at ~12659 eV (denoted as P1) and ~12663 eV (denoted as P2) are observed. The Se K-edge XANES features are mainly due to a direct 1s → 4p dipole transition. The spectrum of CISe-4 (a stoichiometric sample) has a small P1 but a large P2. Conversely, the spectra of CISe-1 (the most Cu-deficient sample) and $In_2Se_3$ have a large P1 but a small P2. Therefore, the P2/P1 intensity ratio appears to be an increasing function of Cu concentration, as shown in Fig. 4d.

It has been reported that P1 corresponds mainly to the In 5sp – Se 4p state, whereas P2 corresponds to Se 4p, which hybridizes with both Cu 4s and In 5sp [19, 47, 48, 51]. Thus, it is reasonable that the overall P2/P1 ratio tended to increase with Cu concentration. However, the drastic P2/P1 increase with increasing Cu concentration cannot be attributed to the presence of certain Cu-containing secondary phases because all samples mostly consist of a single chalcopyrite structure.

The decreasing P1 and increasing P2 intensities can be attributed to changes in the local coordination symmetry of the $Se^{2-}$ ions [19, 47, 50]. In the $CuInSe_2$ structure, $Se^{2-}$ ions are surrounded by two $In^{3+}$ ions and two $Cu^+$ ions forming a tetrahedral coordination (close to a $T_d$ symmetry). Thus, the loss of Cu in the Cu-deficient CISe deformed the tetrahedral coordination, resulting in triangular coordination, as shown in the inset of Fig. 4d.

The FT EXAFS data of the Se K-edge spectra are shown in the inset of Fig. 4c. The peak that appears at R before the phase-shift correction around 2.0–2.6 Å is attributed to the Se-M (M = Cu and/or In) bond. The peak intensity around 2.0–2.6 Å appears to increase gradually with the increase of the Cu concentration. A curve-fitting analysis of the Se-M peak of the FT spectra was performed using ATHENA to obtain structural information about the samples[49]. All samples, except for $In_2Se_3$ reference, were fitted successfully with a two-shell model comprising a short Se-Cu bond and a long Se-In bond. For $In_2Se_3$, a two-shell model with long and short Se-In bonds was employed instead. The sample parameters obtained are listed in Table 2.

**Table 2.** Results of the EXAFS analyses for Se-Cu and Se-In shells. N is the coordination number, $R+\Delta R$ is the phase-shift ($\Delta R$)-corrected interatomic distance, and $\sigma^2$ is the disorder factor. Here, the $S_0^2$ values were assumed to be 0.7 for both shells.

| Sample | Path | N | $R+\Delta R$ (Å) | $\sigma^2$ (Å²) |
|---|---|---|---|---|
| **CISe-1** | Se-In | 2.1 (±0.1) | 2.567 (±0.005) | 0.004 (±0.001) |
|  | Se-Cu | 1.1 (±0.1) | 2.424 (±0.006) | 0.005 (±0.003) |
| **CISe-2** | Se-In | 2.0 (±0.1) | 2.569 (±0.005) | 0.004 (±0.001) |
|  | Se-Cu | 1.6 (±0.1) | 2.414 (±0.005) | 0.005 (±0.002) |
| CISe-3 | Se-In | 2.0 (±0.1) | 2.572 (±0.005) | 0.003 (±0.001) |
|  | Se-Cu | 1.9 (±0.1) | 2.406 (±0.005) | 0.005 (±0.002) |
| CISe-4 | Se-In | 2.0 (±0.1) | 2.571 (±0.005) | 0.003 (±0.001) |
|  | Se-Cu | 2.0 (±0.1) | 2.405 (±0.005) | 0.005 (±0.002) |
| In$_2$Se$_3$-ref. | Se-In (long) | 1.0 (±0.1) | 2.644 (±0.102) | 0.005 (±0.004) |
|  | Se-In (short) | 1.7 (±0.1) | 2.608 (±0.046) | 0.005 (±0.004) |

The coordination number (N) of the Se-Cu shell decreased as the Cu concentration decreased (CISe-4 to CISe-1), whereas the N of the Se-In shell remained almost constant for all the CISe films. This obviously shows that the Se$^{2-}$ ions lose parts of the bondings with the adjacent Cu ions in Cu-deficient samples. Compared to the case of stoichiometric CISe (CISe-4), the Se-Cu bond length in the Cu-deficient samples tended to increase by up to 0.019 Å (for CISe-1), whereas the Se-In bond length decreased only slightly (-0.004 Å). This increase in the Se-Cu bond length is consistent with the results of ab initio calculations for Cu-deficient CISe [16, 28, 51, 52].

**In K-edge XAS.** The In K-edge XANES spectra of the Cu$_x$InSe$_{\sim2}$ films are presented in Fig. 5. The features of the In K-edge XANES spectra reflect the In 1s → 5p electron transition [53, 54]. Overall, the line shapes of the XANES spectra of the four CISe samples are similar. However, the peaks shifted in accordance with the Cu concentration, as shown in the inset of Fig. 5. The absorption edge $E_0$ determined as the energy at the maximum slope, gradually shifted by approximately +1 eV as the Cu concentration decreased. This suggests an increase in the In valence state, which is consistent with the XPS results (Fig. 3b).

**Electronic structure and optical properties.** To reveal the effect of the Cu deficiency on the electronic structure and optical properties, we examined the valence band (VB) features in the XPS spectra. The XPS VB of the Cu$_x$InSe$_{\sim2}$ thin films and the In$_2$Se$_3$ reference (Fig. 6a) exhibited the characteristic features of the valence band in the low-energy region (0–8 eV) of the spectrum. The valence band of the Cu$_x$InSe$_{\sim2}$ thin films consists of the Cu 3d and Se 4p orbital states, whereas that of the In$_2$Se$_3$ reference sample consists of the Se 4p orbital states. The Cu 3d states have relatively sharp features and spin-orbit-split into 3d$_{5/2}$ and 3d$_{3/2}$, but they are near the Fermi level. Meanwhile, the Se 4p orbital states have broad features but mostly maintain BEs higher than ~1 eV. Therefore, we can tell the VB maximum (VBM) is mainly attributed to Cu 3d, whereas it can also be partially attributed to Se 4p [16, 28, 52].

The VBM positions of the thin films were obtained by the linear extrapolation of the leading edge to the extended baseline of the VB spectra. The VBM BEs of CISe-1 to CISe-4 are determined to be 0.37 ±0.02, 0.35 ±0.02, 0.30 ±0.02, and 0.31 ±0.02 eV, respectively. These values indicate that the VBM is lowering slightly as the Cu concentration decreases. It is simply because of the weakening of Cu 3d contribution at the VBM region in accordance with decreasing Cu concentration. In addition, the VBM energies of CISe-1 to CISe-4 are much higher than that of $In_2Se_3$ (BE= 0.92 ±0.05 eV). This clearly shows that all specimens, even for $x$ = 0.34 (CISe-1), have very different electronic structures from those of $In_2Se_3$.

The absorption coefficients in the ultraviolet (UV), visible (Vis), and near-infrared (NIR) light ranges were measured using UV/Vis/NIR optical absorption spectroscopy. Figure 6b shows the absorption coefficient (α) as a function of incident photon energy (hv). The absorption spectra of the $Cu_xInSe_{\sim 2}$ thin films show two absorption edges (A) and (B) at approximately 3.35 eV and 2 eV, respectively. Presumably, feature A may originate from the optical transition from the Cu 3d main peak or broad Se 4p states to unoccupied In 5s, whereas feature B may originate from the transition from the VBM to unoccupied In 5s. The absence of feature B in $In_2Se_3$ supports these assignments because the contribution of Se 4p to the VBM is negligible (Fig. 6a).

Figure 6c shows the Tauc plot ((α hv)$^2$ versus hv) to deduce the direct bandgaps for $CuInSe_2$ and $In_2Se_3$, which are known as direct bandgap semiconductors [20, 22, 23]. The bandgap values ($E_g$) can be deduced from the extrapolation at the highest slopes using the following equation [55]:

$$\alpha h v = A(h v - E_g)^{1/2}, \quad (3)$$

where, A is a constant.

The determined bandgap energies for $Cu_xInSe_{\sim 2}$ thin films with Cu/In ratios of 0.34, 0.46, 0.78, and 0.99 are 1.40, 1.46, 1.51, and 1.51 eV, respectively, while that of the $In_2Se_3$ film is 1.67 eV. The bandgaps of the CISe samples were significantly smaller than those of the $In_2Se_3$ reference. The decrease in the bandgap with decreasing Cu concentration implied that the photon absorption efficiency could be enhanced in the Cu-deficient CISe system. It has been expected theoretically that the Cu deficiency (either Cu vacancy or surplus In) can be "absorption active" so that the relevant defect states can be coupled optically to the conduction bands, resulting in the narrowing of the optical bandgap [16, 17, 28, 34].

**Discussion**

Finding a correlation between composition and optoelectronic structure necessitates a comprehensive knowledge of chemistry, long-range and short-range atomic orders, and electronic structure. In this paper, we addressed those aspects with regard to the impact of the Cu deficiency. The results of the various analyses are summarized in Table 3.

**Composition and chemistry.** The concentrations of each atomic species were deduced by a combination of the EDX analyses of FE-SEM and Cs-TEM measurements. Comparing the [Cu]/[In] concentration ratios before and after selenization (Se vapor deposition) revealed that the cationic ratios were roughly preserved, as intended. Interestingly, the Se ions were uniformly dispersed throughout the film (as shown in the EDX mapping images, Figs. S1† and S2† in Supplementary Information).

Furthermore, there are no noticeable signatures of remnant Se metal in the XRD data (Fig. 2a) or bulk-sensitive Se K-edge XANES spectra (Fig. 4c). This implies that the selenization process was self-limiting, such that only the amount of Se required for full oxidation ($In^{3+}$ and $Cu^+$) was absorbed and spread throughout the film. At the surface, the Se atoms can be oxidized by exposure to air (Fig. 3c). Thus, the fabricated CISe films, apart from the surface region (within a few nanometers in depth), can be regarded as $Cu_xInSe_{\sim2}$.

**Table 3.** Summary of the findings: atomic ratio, crystallinity-related properties (crystallite size and grain size), local structural properties (Se-Cu coordination number and bond length), and electronic structure-related properties (optical bandgap and VBM position).

| Sample | Atomic ratio [Cu]/[In] | Crystallinity-related Crystallite size (nm) | Grain size (nm) | Local structure-related N (Se-Cu) | R+ΔR (Se-Cu) (Å) | Electronic structure-related Bandgap (eV) | VBM (eV) |
|---|---|---|---|---|---|---|---|
| CISe-1 | 0.34 | 10.4 | 44.5 | 1.1 | 2.424 | 1.40 | ~ 0.37 |
| CISe-2 | 0.46 | 11.6 | 52.3 | 1.6 | 2.414 | 1.46 | ~ 0.35 |
| CISe-3 | 0.78 | 13.1 | 59.0 | 1.9 | 2.406 | 1.51 | ~ 0.30 |
| CISe-4 | 0.99 | 15.7 | 71.1 | 2.0 | 2.405 | 1.51 | ~ 0.31 |

**Cu-deficiency effects in structural aspects.** Stoichiometry and microstructure are crucial for determining the crystallinity of the films. Figure 7 shows the average crystallite sizes (from the Debye–Sherrer formula for the XRD peak profiles) and average grain sizes (from FE-SEM) as functions of $x$ and the [Cu]/[In] concentration ratio. Both sizes tended to increase with increasing $x$, suggesting superior crystallinity in the stoichiometric CISe. In Cu-deficient CISe, the Cu deficiency may act as nucleation sites to constitute boundaries between crystallites as to effectively lower the averaged crystallite/grain sizes [37, 56, 57] In addition to the crystallite size, the degree of structural deformation, called microstrain ε, can be quantified using the Stokes–Wilson formula [58, 59]:

$$\varepsilon = \Gamma/(4\tan\theta)\ ,\qquad (4)$$

The ε values are $1.16 \times 10^{-2}$, $8.75 \times 10^{-3}$, $7.22 \times 10^{-3}$, and $6.48 \times 10^{-3}$ for CISe-1, CISe-2, CISe-3, and CISe-4, respectively, and are appended in Fig. 7. The values of the microstrains increased as $x$ decreased (i.e., the system is getting farther from stoichiometry). This suggests that Cu-deficient CISe is structurally less stable than stoichiometric CISe [57, 60, 61]. Nevertheless, Cu deficiency is favorable for low-bandgap optoelectronics because the intrinsic defect states relevant to the Cu vacancy can act as hole-trapping centers in the CuInSe$_2$ system to enhance optical excitation [16, 17, 28, 34].

**Cu-deficiency effect in local structural aspects.** In the chalcopyrite-type CISe crystal structure, the Cu and In atoms are coordinated by four Se atoms, whereas the Se atoms are coordinated by two Cu and two In atoms. In the EXAFS data, features of Cu-Se and In-Se bondings were observed, whereas the signatures of indirect Cu-Cu, In-In, Cu-In, and Se-Se bondings were hardly noticeable. Electron hopping occurred mostly along the networks through the tetrahedral bonding networks of Cu-Se-In-Se-..... Therefore, to understand the electronic structure of the defective CISe system, it is essential to elucidate the coordination under Cu deficiency.

Figure 8a summarizes the coordination numbers deduced from EXAFS analyses of the Se K- and Cu K-edges. The coordination numbers of Cu-Se (3.9) and Se-In (2.0) are maintained very close to those in fully stoichiometric CuInSe$_2$ (4 and 2, respectively), regardless of the Cu concentration. The robust coordination of the remaining Cu atoms is consistent with the predominant Cu$^+$ species, irrespective of the Cu concentration (Fig. 4b), because the local coordination determines the chemical state of the central ion. In contrast, the Se-Cu coordination number drastically reduces from 2.0 to 1.1 with decreasing $x$, which naturally reflects the loss of Cu.

The Se-Cu (Se-In) bond lengths deduced from the Se K-edge EXAFS analyses show a small but systematic increase (decrease) as $x$ ([Cu]/[In] ratio) decreases. For comparison, the Cu-Se bond lengths deduced from the Cu K-edge EXAFS are appended (Table S2† in

Supplementary Information). They are the same within the error bars with the Se-Cu bond length from the Se K-edge EXAFS, confirming the fidelity of the EXAFS processing. The small differences between them may originate from the uncertainty in the chemistry-dependent scattering phase corrections.

Figure 9 shows the Se local structures of the most Cu-deficient CISe (CISe-1) and stoichiometric one (CISe-4). Cu deficiency (either Cu vacancies or $In_{Cu}$ antisites/interstitials) resulted in the elongation (contraction) of Se-Cu (Se-In) bonding. Most plausibly, as one of the $Cu^+$ ions is lost (right panel of Fig. 9), the $Se^{2-}$ ion in the 2In-Se-2Cu coordination moves out of the tetrahedral center being slightly closer to the In ions whereas leaving the remaining $Cu^+$ ion apart. The shorter Se-In bonds could be associated with an increase in the oxidation number of $In^{3+}$, as shown in Fig. 3b and Fig. 5. This is because the increased charge number in $In^{3+\delta}$ would tend to attract the anions slightly more. The influence of the local structural evolution upon the Cu deficiency is discussed in the next section.

**Cu-deficiency effect in optoelectronic aspects.** In $CuInSe_2$, In-Se ionic bonding is stronger than Cu-Se bonding mainly because the In ions have a large positive charge (+3) compared to the case of $Cu^+$ ions so that the attractive interaction in In-Se is stronger than that in Cu-Se [16, 17, 28, 52, 62].

The middle panel in Fig. 9 depicts the band structure of CISe-1 and CISe-4 with the energy values estimated by XPS and the optical spectroscopy (Fig. 6). In 5s CB lowering associated with local structural deformation shown in Fig. 9 may play a key role in the bandgap narrowing.

In the case of Cu-deficient CISe, the partial loss of attractive Cu-Se interaction will reduce the lengths of the remaining In-Se bonds, and consequently the In valence will increase to $3 + \delta$ because of the Coulomb energy gain due to increased ionization number in the strong ionic bonds. Since the $In^{3+\delta}$ valence shell in Cu-deficient CISe will accept the photoexcited electron more easily than the $In^{3+}$ valence shell in stoichiometric CISe, the energy cost to excite a VB electron to the CB will become smaller. Therefore, the CBM, which is mainly contributed by the In 5s orbital states, will be lowered as the Cu concentration (*x*) decreases. Meanwhile, the VBM is mostly contributed by fully filled Cu 3d states (with a partial contribution of Se 4p); therefore, it is rather insensitive to the detailed evolution of the charge distribution with increasing *x*. Therefore, the VBM energies would remain roughly constant at *x*.

In contrast, the bandgap, which is the energy difference between the CBM and VBM, decreases as the CBM is lowered. Figure 8c shows the bandgap values and VBM energies as functions of *x*. Indeed, the VBM is lowered but only slightly (~0.06 eV) with decreasing *x*, owing to the weakened Cu 3d contribution (Fig. 6a), while the bandgap decreases by 0.11 eV. This suggests that the weakening of the Se 4p–Cu 3d interband or formation of excitonic acceptor levels has a limited influence on the bandgap change [16, 17, 28, 34, 52, 62].

Meanwhile, the CBM is lowered by ~0.17 eV with decreasing $x$. Accordingly, the bandgap is reduced from 1.51 eV (CISe-4) to 1.40 eV (CISe-1), which is optimal for the photon absorber in solar cells [63-65], and this reduction is mostly contributed by the lowering of the In 5s CBM. The lowered CBM is associated with an increase of In valence (as is confirmed by XANES in Fig. 5) and at the same time with the off-center displacement of the Se ion out of the Cu-Se-In tetrahedral networks (EXAFS in Fig. 4) under the Cu-deficient condition. This implies that the electronic structure can be tuned by controlling detailed chemistry and local structure.

Therefore, it can be concluded that the electronic structure (VBM and CBM) and the optoelectronic properties (bandgap) can be engineered by the Cu-deficiency control while maintaining the chalcopyrite crystal structure (Cs-TEM and XRD). The local structural study (EXAFS) revealed the details of the structural reconstruction in the Cu-Se-In network upon the loss of Cu, which might stabilize the chalcopyrite structure without any phase segregation, even when approximately half of the Cu is missing (CISe-1).

The sufficient supply and self-limiting permeation of Se vapor in the selenization process might be crucial for stabilizing the excessively ionized state of $In^{3+\delta}$ and the lowered bandgap; otherwise, parts of the cations (In or Cu) would be reduced to form metallic clusters, which was not the case in this work. Thus, efforts should be made to optimize the optoelectronic properties by controlling the synthesis parameters.

**Conclusions**

$Cu_xInSe_{\sim 2}$ thin films with $x$ = 0.34, 0.46, 0.78, and 0.99 prepared by consecutive e-beam evaporation and Se vaporization processes exhibited excellent chalcopyrite crystal structure tolerance in Cu-deficient conditions with a superior optoelectronic property with a bandgap of 1.4 eV. Various chemical and structural investigations, including XRD, FE-SEM, Cs-TEM, UV/Vis/NIR, XPS, XAS, and EXAFS analyses, demonstrated that the Cu deficiency can be stabilized with accompanying local structural evolution (bond length changes in Cu-Se-In networks) while maintaining the $CuInSe_2$ crystal structure and that the resultant orbital states $In^{3+\delta}$ 5s reconstruction is mainly responsible for the bandgap narrowing. The facile tunability of the electronic structure with chemical and structural stabilization suggests that Cu-deficient $Cu_xInSe_{\sim 2}$ is a promising nontoxic optoelectronic material for photovoltaic devices.

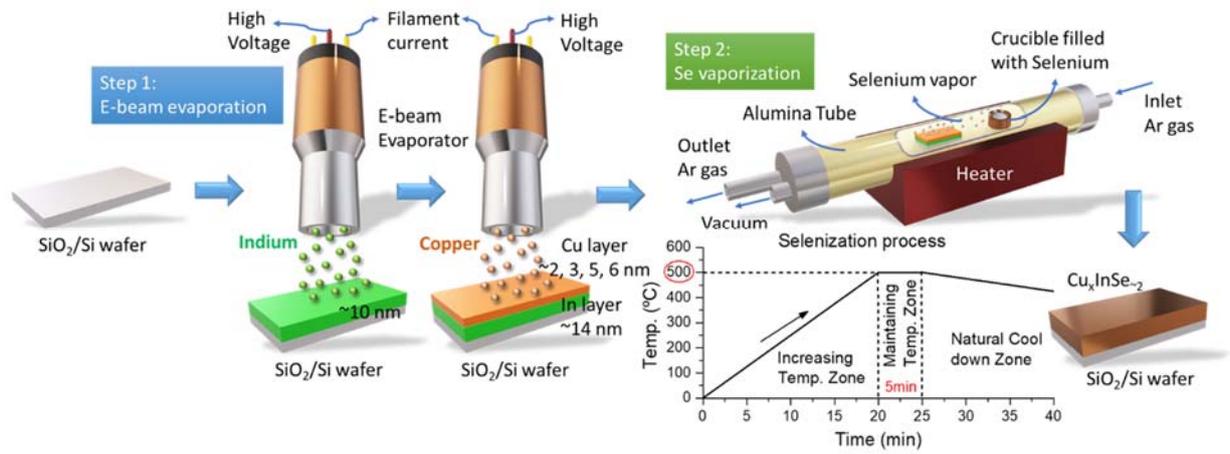

**Figure 1.** Schematics of the two-step preparation method for $Cu_xInSe_{~2}$ thin films: The first step is the e-beam evaporation of In and Cu, and the second step is the Se vapor deposition process in a tube furnace.

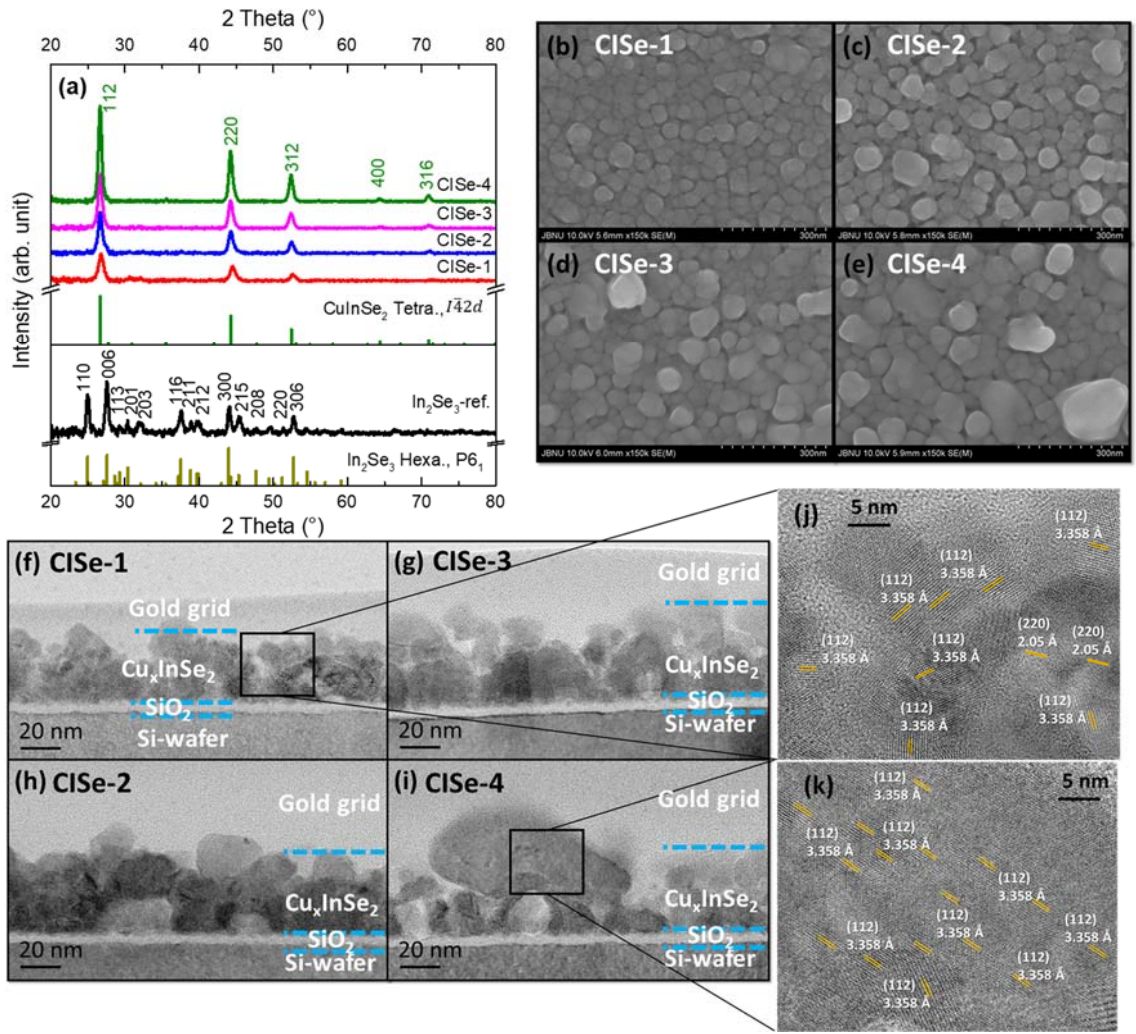

**Figure 2.** (a) XRD patterns of $Cu_xInSe_{\sim 2}$ thin films ($x$ = 0.34, 0.46, 0.78, and 0.99) and a reference $In_2Se_3$ film, and the reference patterns from the database (bar-typed). (b-e) FE-SEM images and (f-i) Cross-sectional TEM images of the respective $Cu_xInSe_{\sim 2}$ films. (j,k) The enlarged views for $x$ = 0.34 and $x$ =0.99, show the same d-spacings of 3.358 Å for (112) and 2.05 Å for (220) reflections, suggesting uniform and identical crystal structure regardless of $x$.

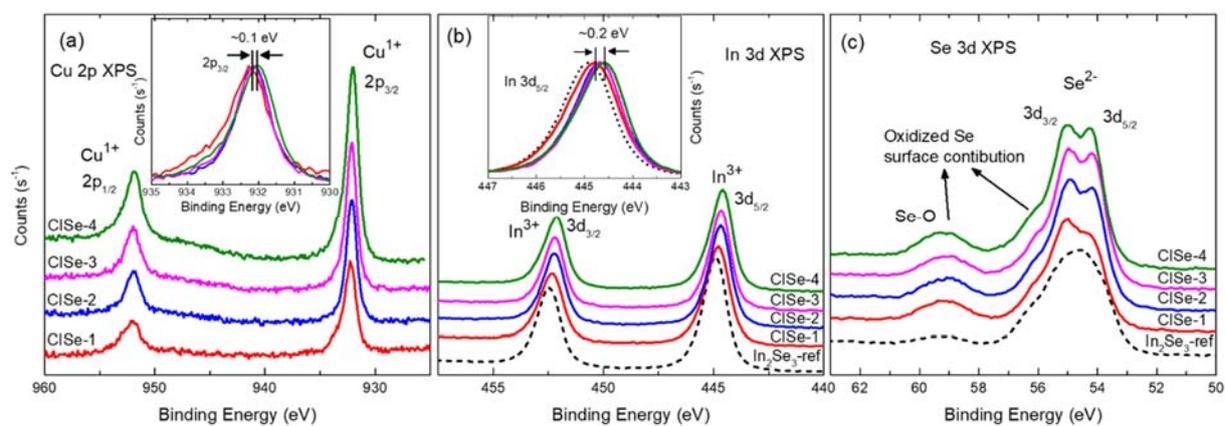

**Figure 3.** (a) Cu 2p, (b) In 3d, and (c) Se 3d core-level XPS spectra of the $Cu_xInSe_{\sim2}$ thin films at $x$ = 0.34 (CISe-1), 0.46 (CISe-2), 0.78 (CISe-3), and 0.99 (CISe-4) and the $In_2Se_3$ reference.

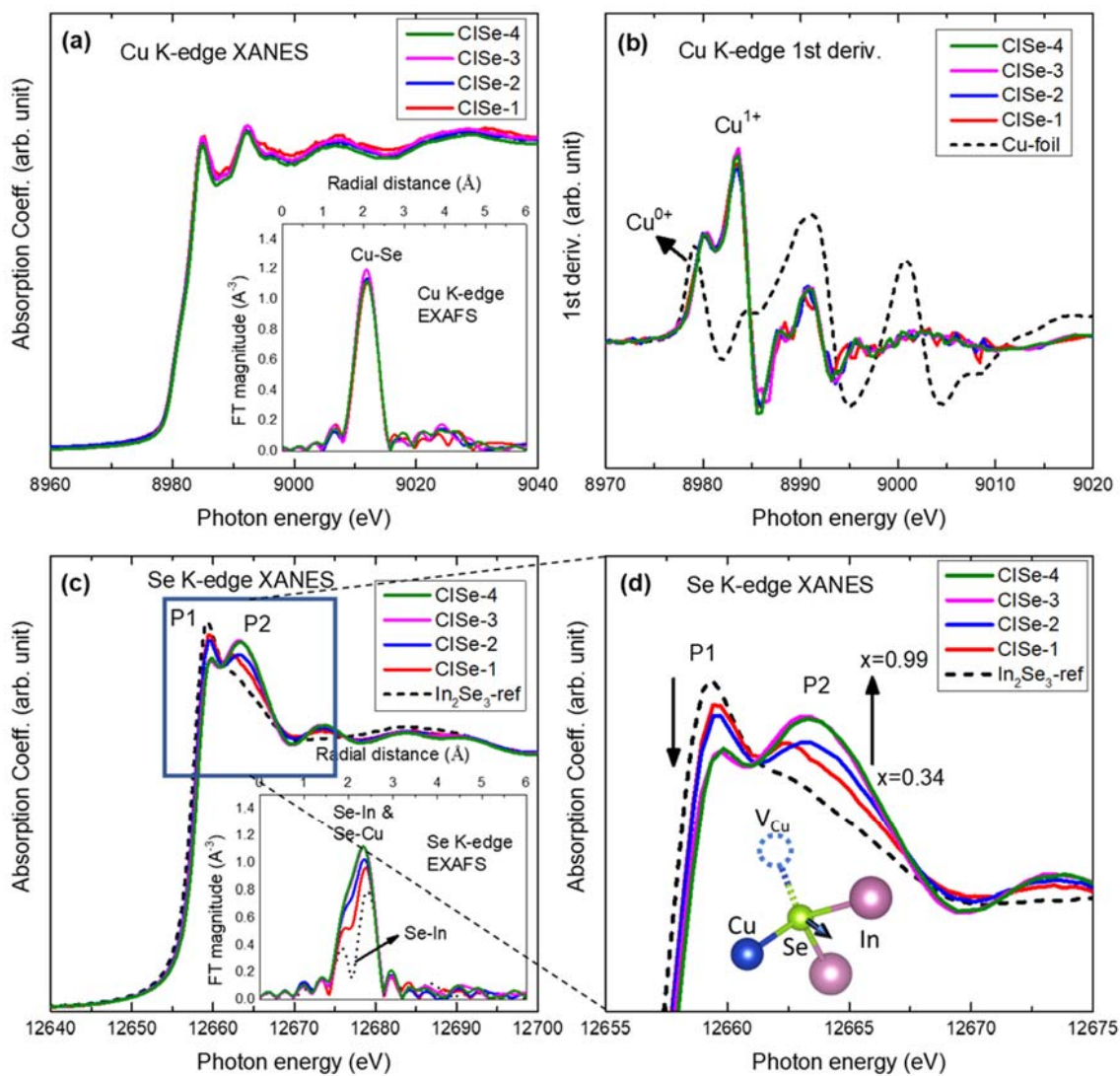

**Figure 4.** (a) Cu K-edge XANES of the 4 CISe samples and (b) their first-derivative spectra together with the spectrum of a Cu foil. (c) Se K-edge XANESs of the $Cu_xInSe_{\sim2}$ samples and $In_2Se_3$ reference, and (d) the enlarged view in the white-line energy range. The FT spectra of the Cu K- and Se K-edge EXAFSs were appended in the insets of (a) and (c), respectively.

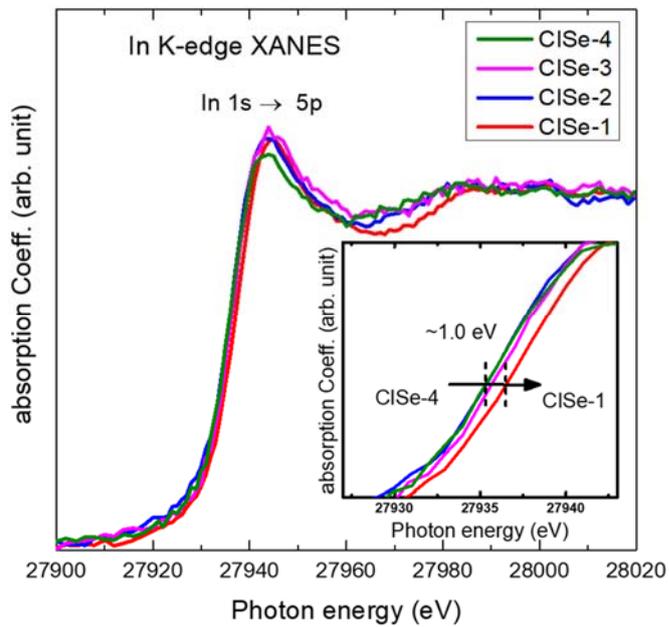

**Figure 5.** In K-edge XANES spectra of the 4 CISe films. The In K-edge XANES spectra for all the samples exhibited a similar white line with a slightly higher-energy shift (~1.0 eV) as the Cu concentration decreases.

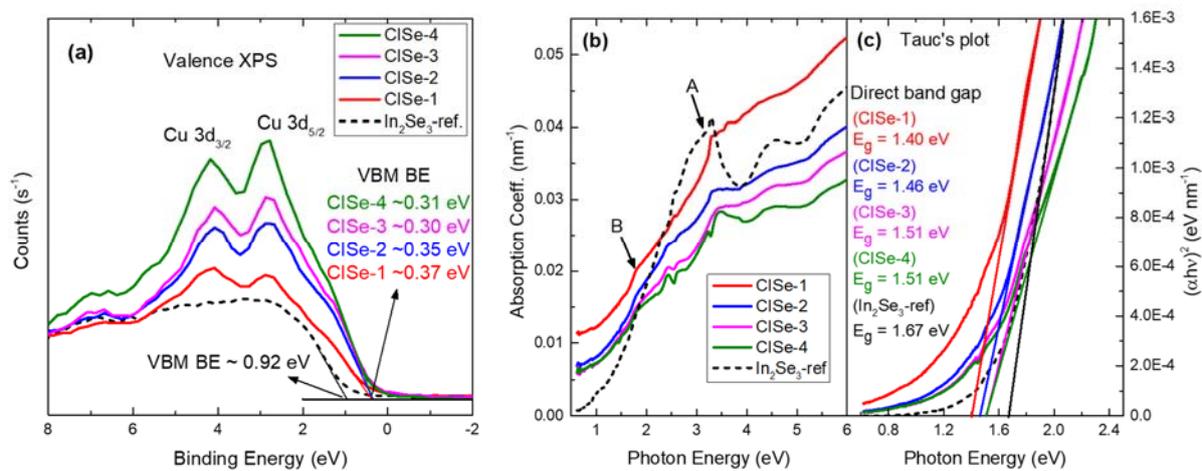

**Figure 6.** (a) The XPS VB spectra and the (b) UV/Vis/NIR absorption spectra with the (c) Tauc plot for direct bandgap calculations.

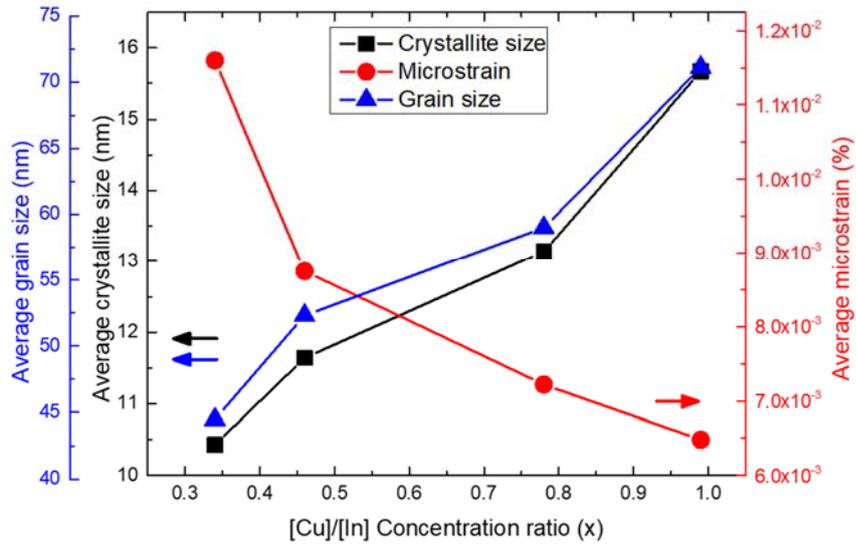

**Figure 7.** Morphology and microstrain as a function of [Cu]/[In] concentration ratio. Both sizes tend to increase as $x$ increases, meanwhile, the microstrains decrease as $x$ increases, indicating superior crystallinity in the stoichiometric CISe.

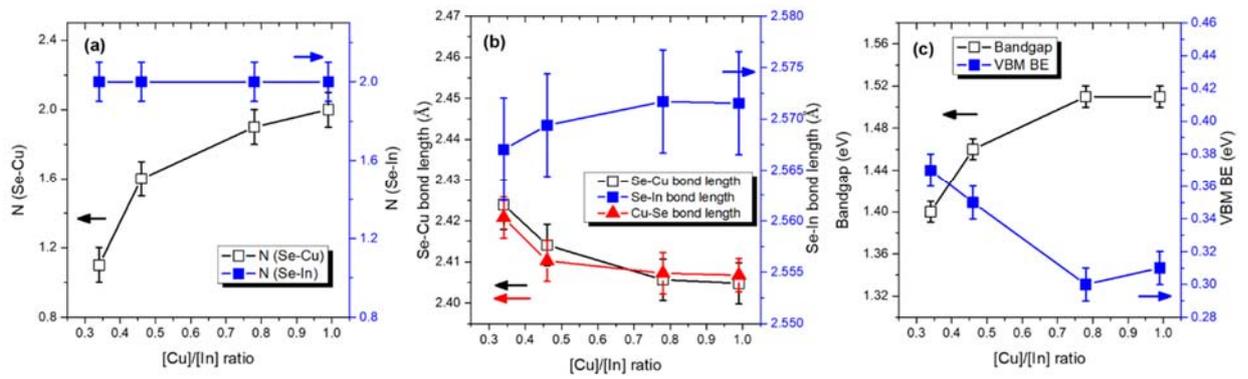

**Figure 8.** Local structural information ((a) numbers of nearest neighbors and (b) the bond lengths) and (c) the electronic structure information (bandgap and VBM energy) as functions of [Cu]/[In] concentration ratio.

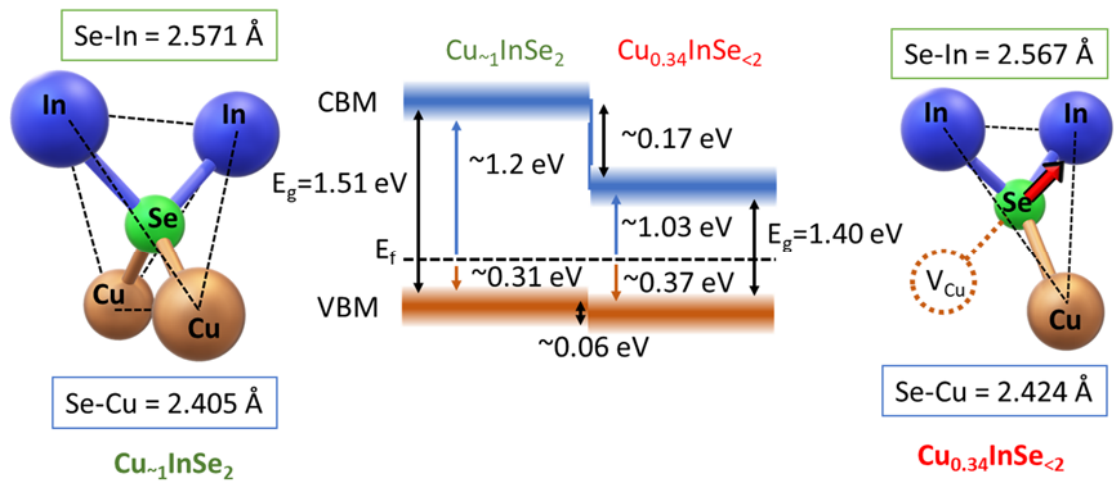

**Figure 9.** Local structural evolutions from stoichiometric $Cu_{\sim1}InSe_2$ (left panel) to Cu-deficient $Cu_{0.34}InSe_{<2}$ (right panel), and the difference in band structure between them (middle panel).